# FEEDBACK SYSTEMS FOR FCC-EE

Alessandro Drago
Istituto Nazionale di Fisica Nucleare, Laboratori Nazionali di Frascati
Via Enrico Fermi 40, Frascati, Italy

*Abstract*

In this paper, some preliminary considerations on the feedback systems for FCC-ee are developed. Bunch-by-bunch feedback systems have been designed in the last years for other e+/e- colliders like PEP-II, KEKB, DAFNE, SuperB and SuperKEKB. In all these cases, similar approaches have been implemented, even if some design variations have been suitable or necessary for different reasons. Bunch-by-bunch feedback systems are based on the concept that the barycenter of each bunch moves with harmonic motion around the equilibrium point in three planes (L, H, V). The feedback copes with the forcing excitation by producing damping correction for each individual bunch. This is possible managing every single bunch by a dedicated processing channel in real time. For FCC-ee the very high number of stored bunches requires much more power in terms of processing capability for the feedback systems. Ring length (100 Km) and very low fractional tunes must be also considered requiring for a more effective strategy in the feedback system design.

## INTRODUCTION

A new tunnel with a circumference of 100 Km around the CERN area could host the proposed FCC-ee (Future Circular Collider e-/e+), in the past also called TLEP (Triple LEP). As alternative, the tunnel can host FCC-hh, a 100 TeV center-of-mass energy-frontier hadron-hadron collider or FCC-he, a proton-electron collider. Beside the lepton accelerator, the hadron one faces additional challenges, such as high-field magnet design, machine protection and effective handling of large synchrotron radiation power in a superconducting machine [1] [2].

FCC-ee [3][4], operating at four different energies for precision physics of Z, W, and Higgs boson and top quark, represents a significant push in terms of technology and design parameters. Pertinent R&D efforts include the RF system, top-up injection scheme, optics design for arcs and final focus, effects of beamstrahlung, beam polarization, energy calibration, and power consumption. Finally, feedback systems in the three oscillation planes (H, V, L) are necessary, and some preliminary considerations are carried on in this paper.

To achieve this goal, in the following a fast sketch of the foreseen instabilities impacting the possible design of the FCC-ee feedback systems is reported.

In the past two decades, bunch-by-bunch feedback systems have been designed for several e+/e- colliders like PEP-II [5] [6] [7], DAFNE [8], KEKB [9] and more recently for SuperB [10] (feedback built and installed at DAFNE [11]) and SuperKEKB [12].

In all these cases, very similar or identical approaches have been implemented, even if some design variations have been possible for technological progress or convenient for specific reasons.

All these feedback systems are based on the concept that the barycenter of each bunch moves with harmonic motion around the equilibrium point in each of the three planes (H, V, L).

The feedback copes with the forcing excitation by calculating individual damping correction kicks for each bunch. This is possible by managing in real time every single bunch by a dedicated processing channel implementing a FIR (Finite Impulse Response) filter at n taps, with n from 1 to 32, chosen by the operator. In each system, the phase response must be found experimentally with great care to give the best correction kick for each bunch. Betatron and synchrotron phase advance at pickups and kickers determines the filter setup along with other parameters.

For the FCC-ee feedback systems, a similar design is proposed in this paper taking also in consideration the peculiarities of the collider: a very high number of stored bunches and a huge harmonic number, fast instability growth rates and remarkable ring length.

Figure 1: Parameter list from FCC Week 2015 M.Migliorati's talk with arrows indicating the most relevant parameters for a feedback design point of view.

**Parameter list - FCC-ee Z-pole, crab waist, 2 IPs**

| parameter | Z | W | H | t |
|---|---|---|---|---|
| Circumference (km) | 100 | 100 | 100 | 100 |
| Beam energy (GeV) | 45.5 | 80 | 120 | 175 |
| Beam current (mA) | 1450 | 152 | 30 | 6.6 |
| RF frequency (MHz) | 400 | 400 | 400 | 400 |
| RF Voltage (GV) | 0.2 | 0.8 | 3 | 10 |
| Mom compaction [$10^{-5}$] | 0.7 | 0.7 | 0.7 | 0.7 |
| Bunch length [mm](*) | 1.63 | 1.98 | 2.0 | 2.1 |
| Energy spread(*) | $3.7 \times 10^{-4}$ | $6.5 \times 10^{-4}$ | $1.0 \times 10^{-3}$ | $1.4 \times 10^{-3}$ |
| Synchrotron tune | 0.025 | 0.037 | 0.056 | 0.075 |
| Bunches/beam | 90300 | 5162 | 770 | 78 |
| Bunch population [$10^{11}$] | 0.33 | 0.6 | 0.8 | 1.7 |
| Betatron tune | 350 | 350 | 350 | 350 |

(*) without beamstrahlung (no collision)

## FORESEEN INSTABILITIES

The preliminary evaluation of the foreseen instabilities is based on the talks held by M. Migliorati at the FCC Week 2015 (First Annual Meeting of the Future Circular Collider Study) [13] and in a March 2016 CERN meeting (First Annual Meeting of the Future Circular Collider Study) [14]. In the following Fig. 1 a parameter list is shown with the arrows indicating the most relevant parameters for a feedback design point of view.

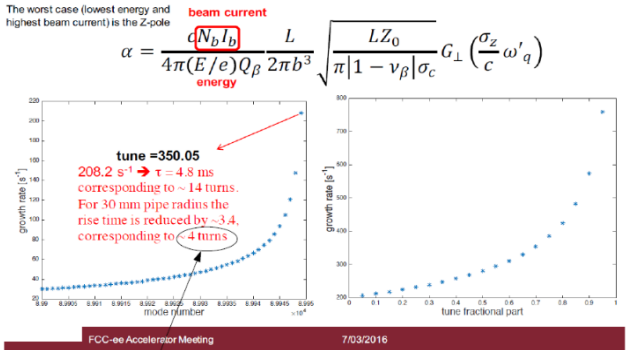

Figure 2: RW instability growth rate is evaluated in four turns (from M. Migliorati's talk during the March 2016 CERN meeting). Fractional tune is 0.05.

The harmonic number H is 133600. Usually the bunch-by-bunch feedback systems process all the buckets without selecting the full and the empty ones. This is a design choice that simplifies and makes faster the data processing.

A preliminary evaluation of RW (resistive wall) instability growth rate has been discussed in the March 2016 meeting and one of the results is shown in the Fig. 2.

From the transverse point of view the most important instability is due to RW. The value computed from the impedance model gives a growth rate of 4 turns, that is an extremely fast value from a feedback performance point of view. Nevertheless, if we consider the e-cloud effects expected in the positron ring, the growth rates could be even much worst. Furthermore, the fractional tune is 0.05, that is a very low frequency.

If we consider the DAFNE case, the resistive wall instability is faster in the e+ ring than in the e- ring by a factor 10. The more relevant instability effect is in the horizontal plane [15] [16].

## FEEDBACK BASICS

To introduce a possible design, the feedback system description is organized in several main blocks as shown in the Fig. 3.

The blocks are: pickups, analog front-end, DPU (digital processing unit), including analog to digital converter working with at least 12 bits, better if 14 or 16, multiplexer, FIR filter for each bunch, demultiplexer, digital to analog converter, analog back end, timing including also delay lines, power amplifiers, kickers, and operator interface for remote control.

Analyzing more in depth every block and starting from the **pickups**, there are no special requirements, apart a very good H/V beta to increase the signal to noise ratio.

In the transverse plane, usually this condition requires to use different BPM (beam position monitor) as feedback horizontal and vertical pickups.

The **kickers** are different for the two cases: cavity type kicker [17] should be implemented for the longitudinal plane while stripline type kickers for transverse planes. For both cavity and stripline kicker, the impedance needs to be carefully evaluated. As in the previous e+/e- colliders, a high beta (H/V) is required for the transverse kickers to have the best feedback performance. Implementing separate kickers for the horizontal and the vertical case helps to have a high beta in both the cases.

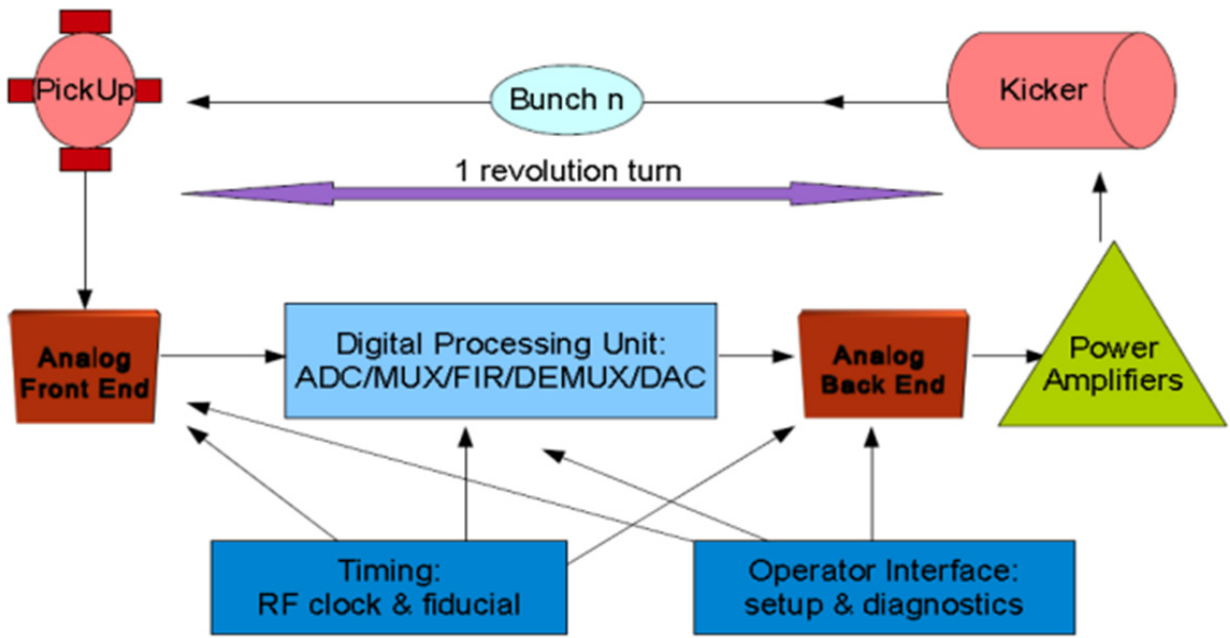

Figure 3: Feedback main blocks. They can differ if transverse or longitudinal system.

Both kinds of kicker should assure adequate performance with signals having very low frequencies, even of the order of one Hertz. This is because the stripline kicker has a bandwidth starting from dc and the cavity kicker requires a signal modulated at 1-1.5 G Hertz, allowing to manage even dc voltage.

The **Analog Front End** differs partially from the longitudinal and the transverse cases. Basically, it is designed around custom PCB comb filters at 4*RF (H, V) or 6*RF (L). The same scheme has been implemented for SuperB (installed at DAFNE) and SuperKEKB. In both cases the modules have been assembled in-house.

Regarding the **DPU** and considering the complexity of this block, it will be convenient to implement the same type of unit for all the feedback types (H, V, L) to reduce design and fabrication efforts. There is a warning about the very high number of processing channels. This point will be discussed in the next paragraph. Note that in the DPU the signal propagation delay is of the order of 600 ns. The DPU has the goal to implement the FIR filters that compute the correction signal for each bunch.

The **Analog Back End** for the transverse feedback systems will work in base band, so usually only a level adapter (pre-amplifier stage with splitters) is necessary. For the longitudinal feedback, as in the previous designs, a double modulation scheme is proposed, both AM and QPSK. To make effective these modulations on each bunch, a precise analog delay line with high bandwidth must be included. In both cases, transverse and longitudinal, the modules can be assembled in-house.

About the **power amplifiers**, there are models with an adequate bandwidth that of course will be different for transverse and longitudinal cases. Typically, they have between 250W and 2kW power and they are commercially available (though very expensive). The pulse response must be evaluated to avoid cross-talk between adjacent bunch. Note that the performance at very low frequency needs to be checked to manage correctly the bandwidth.

**Timing** specifications are like for the other colliders considered previously, that means a jitter within 10 ps. However, the transmission over such distances, of the order of tens of Km, needs to be managed by an adequate technology mainly to maintain jitters in the specification.

The **operator interface** takes care of the correct running of the systems, the real time and off line diagnostics, the verification and the implementation of the best setup. In general, the setup will be different for each individual feedback system and it must be checked with single and multiple bunches.

## The First Critical Point

As said above, the number of bunches is very high, furthermore the bunch-by-bunch feedback systems currently implemented do not use a lookup table to select the filled and the empty buckets, so they must process all the H buckets (H=harmonic number=133600) even if empty. Changing this strategy in the design can be feasible but not necessarily convenient from a design point of view.

By the way, at the present the SuperKEKB feedback processes 5120 bunches, that is the highest harmonic number for currently operative lepton colliders.

Note that the damping ring of ILC, still not approved, should have H~=7000 and CepC proposal should have H=118800 but with few bunches.

In conclusion for FCC-ee, each feedback system needs a processing power such as 133600 / 5120 = 26 times the SuperKEKB systems. An advantage is the much slower revolution frequency of FCC-ee that compensates partially the high number of processing channels.

Concluding, the DPU design is not trivial, requiring an extremely high computing power that must be implemented by custom modules based mostly on FPGA technology.

Luckily the FPGA technology is growing fast, so the goal is demanding but feasible. Of course, a new DPU design is necessary.

## *The Second Critical Point, Fb Damping Capability*

It is a common point of view and an experimental result too that a bunch-by-bunch feedback in e+/e- collider can damp the instabilities up to 10 revolution turns, even if, at very low fractional tunes, this result ought to be checked. However, reporting this evaluation in terms of revolution turns helps to correlate the feedback performance for very different length accelerator rings.

Anyway, if the oscillation frequency is very low, for example with fractional tune < 0.09, difficulties can arise to damp very fast instabilities in only 10 revolution turns.

The "standard" performance can be achieved by installing one feedback system for relatively high beam currents (1-3A) with an amplifier section of the order of 1 or 2 kW total power.

The limit to increase the feedback gain is basically due to the noise present in the loop that in large part comes outside the system, from the pickup, and only in small part from inside the feedback. In this case it is due to analog electronics and quantization noise.

## *DAFNE 2008 Experiment*

About the previous topic in the year 2008 at DAFNE the author did an experiment by implementing two separate feedback systems working in the same plane (horizontal e+) [18].

This was necessary for damping a very fast horizontal mode induced by e-clouds and waiting for the fabrication of a new feedback stripline kicker with a much better shunt impedance.

Having the availability of a second kicker in the ring (see Fig. 4) installed to be used as test injection kicker, a new transverse horizontal feedback was put in operation in the positron ring. To avoid managing too complicate timing setup including also difficulties to check the feedback performance, two complete loops were implemented including a second pickup and another DPU. A big advantage of the double feedback strategy is the capability to check easily and separately the correct setup and performance of each system by turning off the other one.

The "trivial" result obtained was that the damping times of the two feedback systems add up about linearly within the measurement error (10-20%). Growth and damping rate are computed by fitting routines running off-line to analyze the stored bunch-by-bunch data. An example of this data analysis done in the 2008 runs is shown in the Fig. 5.

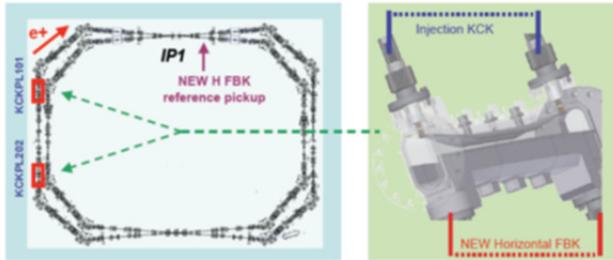

Figure 4: The injection kicker used as horizontal feedback kicker (on the right) and the placements in the e+ ring (on the left).

The modal instability growth rates for the positron beam were 34.5 ms-1 at 560 mA and 712 ms-1 at 712 mA of beam current.

The damping rates measured for mode 119 (=120-1 that is -1 mode as computed by the e-cloud DAFNE model) are also shown in the same fig. 5 and were:

-127 ms-1 for a single feedback loop, that means damping in 24 revolution turns (=7.8 microseconds)

-233 ms-1 by using two cooperating feedback systems that means damping in 13 revolution turns (= 4.3 microseconds).

Note that in DAFNE the harmonic number is 120 and one revolution period (turn) is 324 ns. It should be remembered also that having a DPU propagation delay of 400-600 ns makes impossible to kick the bunch after just one turn and two turns are necessary. This gives a little loss of performance evaluated in about 15%. In conclusion, there is a clear experimental demonstration that the damping rate of the feedback has been doubled by doubling systems and power. Nevertheless, a simulation model could clarify better range and performance of the linearity of the behavior. More important is that two feedback systems in the same ring and plane have worked very well together cooperating perfectly without loss of power. Furthermore, during May 2016, the double feedback technique has been implemented again and tested with same results at SuperKEKB by M. Tobiyama confirming the approach validity.

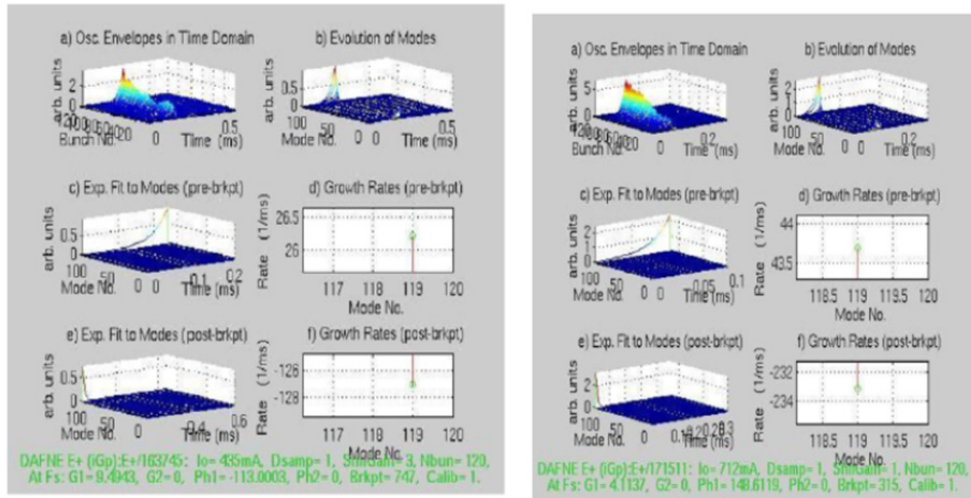

Figure 5: Growth rates and damping rate measured at DAFNE in the year 2008 in the e+ horizontal plane by using one and two feedback systems.

It must be noted that having more kickers placed in different parts of the ring requires for a complete duplicate of the feedback to simplify the setup because both timing and phase response of the systems are different.

In the previous case (year 2008) the systems used 2x250W amplifiers for each feedback. After implementing a new kicker with a much better shunt impedance, there was no more need of the second system.

## FCC-EE FEEDBACK PROPOSAL

In order to perform an effective feedback scheme able to give solution to the FCC-ee beam dynamics problems as outlined in the second chapter, a multiple and distributed feedback approach is proposed. The scheme can be implemented in two different proposals, the first one by maintaining the usual feedback scheme even if updated, the second one by implementing an innovative and more complicate design of DPU to apply correction signal by shortening the loop delay.

A question could arise: why do not implement only one feedback loop with a very high gain by using a big number of power amplifiers to have a faster damping time?

The answer, based on the experience, is: because the noise entering in the pickup cannot be filtered completely and increasing gain and power makes an enlargement effect of the bunches, that could become very evident in the vertical size, and could also push feedback to performance saturation.

### Proposal 1: Cooperating Feedback Systems

Considering a damping time of about 10 turns for each feedback system (feasibility already demonstrated in other lepton colliders with tune > 0.09), it will be necessary to implement more feedback stations, most likely between 4 and 6.

There are some drawbacks in this strategy because a larger number of kickers and pickups increases the ring impedance. Another minor drawback is in the more complicated timing and setup operations.

An important advantage will be to have the possibility to apply correction kicks distributed along the ring. A second advantage is the following: by implementing this strategy it could be possible to achieve, if necessary, the theoretical damping limit of 1 revolution turn by installing more feedback stations, maybe by implementing 10-12 loops. As said previously very low tune frequencies can lower the feedback performance in any case.

Evaluating this scheme, it is obvious that a damping rate faster than 1 turn cannot be not achieved. This is because the correction kick can be applied only with 1 turn delay after processing the acquired signal. The DPU propagation delay cannot allow to kick faster. In the Fig. 6 an example of cooperating feedback scheme is shown. It is implemented by 4 stations placed along the 100 Km long FCC ring. The foreseen damping rate is 2.5 revolution turns.

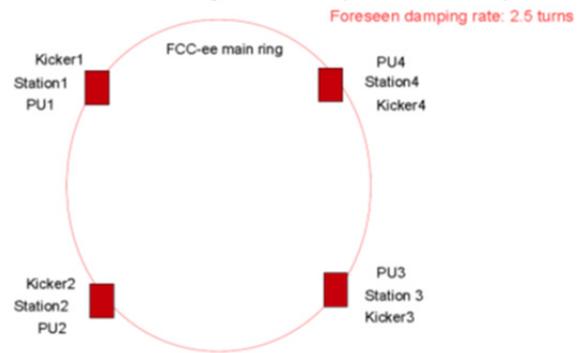

Figure 6: Example of cooperating feedback scheme implemented by 4 stations placed along the 100 Km long FCC ring. The foreseen damping rate is 2.5 revolution turns.

### Proposal 2: Cooperating And "Feeding Forward" Systems

Of course, the ring length is a critical point for managing the feedback in terms of timing and control of the correct performance of the system. Nevertheless, a 100 Km ring length is also a very interesting opportunity to design an innovative scheme of feedback system [19] that is not convenient and neither possible in shorter rings. The idea is to make shorter than 1 turn the correction signal path. In this way, it will be feasible to achieve damping rates even faster than 1 revolution turn.

The "feeding forward" approach will somewhat change the usual system scheme, however not as much as it would seem. The phase response will be controlled by individual bunch FIR filter inside the DPU in this case too.

The implementation would be a big challenge from a technological point of view: it will be necessary to send the correction signal in a way to arrive to the kicker location before the arrival of the bunch that must be corrected. As example, few cases are analyzed below.

First, a "feeding forward" system can be designed that takes the input signal at the pickup 1 (PU1) and, after processing it in the station 1, sends the correction signal to the station 2 where kicker and power amplifier section are placed. The correction signal path follows the main ring diameter and hence it is 31.8 km long. The foreseen damping rate is 5 turns, the half of the standard approach.

In the Fig. 7 the "feeding forward" system is implemented by two stations and two loops. The system 1 takes the input signal at the pickup 1 (PU1) and, after processing in the station 1, sends the correction signal at the station 2 where kicker 1 and power section 1 are placed. Vice versa the system 2 takes the input signal at the pickup 2 (PU2) and, after processing it in the station 2, sends the correction signal to the station 1 where kicker 2 and power amplifier section 2 are placed. The correction signal path is along the main ring diameter and it is 31.8 km long. The foreseen damping rate is 2.5 turns.

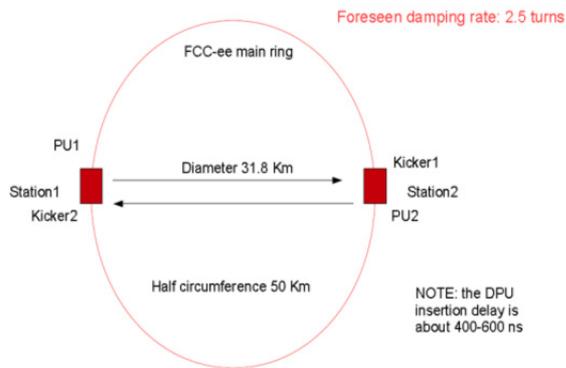

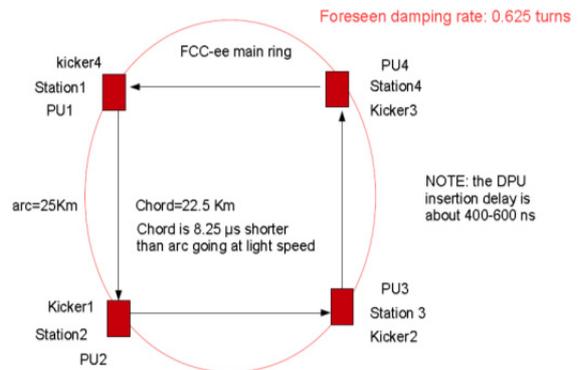

to the station 2 where kicker 1 and power section 1 are placed.

Figure 7: "Feeding forward" system implemented by two stations and two loops. The system 1 takes the input signal at the pickup 1 (PU1) and, after processing it in the station 1, sends the correction signal to the station 2 where kicker 1 and power section 1 are placed. Vice versa the system 2 takes the input signal at the pickup 2 (PU2) and, after processing it in the station 2, sends the correction signal to the station 1 where kicker 2 and power amplifier section 2 are placed. The correction signal path is on the main ring diameter and it is 31.8 km long. The foreseen damping rate is 2.5 turns.

In the Fig. 8 a "feeding forward" system is implemented by one station and one loop. The system 1 takes the input signal at the pickup 1 (PU1) and, after processing it in the station 1, sends the correction signal to the station 2 where kicker 1 and power section 1 are placed. The correction signal path is on the chord related to a quarter of circumference arc and it is 22.5 km long. The foreseen damping rate is 2.5 turns.

Figure 9: "Feeding forward" system implemented by four stations and four loops. The system 1 takes the input signal at the pickup 1 (PU1) and, after processing it in the station 1, sends the correction signal to the station 2 where kicker 1 and power section 1 are placed. The correction signal path is on the chord related to a quarter of circumference arc and it is 22.5 km long. The systems 2, 3 and 4 similarly take the input signals at the PU2, PU3, PU4 and after processing them in the station 2, 3 and 4 they send the correction signal to the station 3, 4 and 1 where kickers and power sections are placed. The foreseen damping rate is 0.625 turns.

The correction signal path is on the chord related to a quarter of circumference arc and it is 22.5 km long. The system 2, 3 and 4 similarly take the input signals at the PU2, PU3, PU4 and, after processing them in the station 2, 3 and 4, they send the correction signals to the station 3, 4 and 1 where kickers and power sections are placed. The foreseen damping rate in this case is 0.625 turns breaking the 1 turn limit with only four systems.

### How to Transmit the Correction Signal

Efficient correction data transmission is not a simple task. A very preliminary idea could be to transmit correction signals by using lasers, with transmission lines in the vacuum, but the alignment, for distances of 22 or 32 km, is not easy. This technique seems also very expensive.

In theory the chord length could be shorter than 22 km, but the arc path should be long enough to recover the DPU insertion delay of 400-600 ns plus cable delays.

Radio-frequency transmission can also be considered [20] but it should be evaluated if it will be affected by noises caused by bad weather or of other natural or artificial origin.

A simpler solution seems to be the transmission of the digital data by optical fibers, compacting the correction signals by an efficient code, maybe treating them in blocks. In this way, the data should be applied to the kicker before the arrival of each bunch that needs to be corrected. A study on the possible codes should start taking as example the well-known MP3 coding.

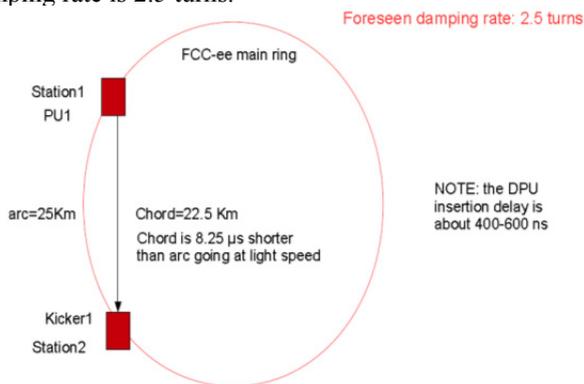

Figure 8: "Feeding forward" system implemented by one station and one loop. The system 1 takes the input signal at the pickup 1 (PU1) and, after processing it in the station 1, sends the correction signal to the station 2 where kicker 1 and power section 1 are placed. The correction signal path is on the chord related to a quarter of circumference arc and it is 22.5 km long. The foreseen damping rate is 2.5 turns.

In the Fig. 9 the "feeding forward" system is implemented by four stations and four loops. The system 1 takes the input signal at the pickup 1 (PU1) and, after processing it in the station 1, sends the correction signal

## CONCLUSION

The feedback systems for FCC-ee can be based on the designs developed for other previous e+/e- colliders as PEP-II, KEK, DAFNE, SuperB and SuperKEKB. The same DPU (digital processing unit) can be used for longitudinal and transverse systems, while analog modules and kickers need to be different in the transverse or longitudinal cases.

A DPU managing more than 100k separated bunch/bucket signals is feasible but it requires efforts for redesigning the present systems in a more compact design.

By implementing multiple cooperative feedback systems and maintaining the "traditional" design scheme it will be possible to damp up to 1 revolution turn, considering fractional tunes > 0.09. This approach has been tested at DAFNE in the year 2008 with very good results. For lower fractional tunes some R&D is necessary.

Damping in less than one revolution turn is possible only changing the usual feedback strategy and implementing an innovative "feeding forward" strategy.

This new approach can be implemented because of course a chord or the diameter are shorter than the related arc and in this way, it is possible to compensate the DPU insertion delay (400-600 ns). Note that this scheme is feasible and convenient only for very long accelerator rings.

A "feeding forward" system, very challenging to implement, requires strong technological efforts to modify (partially) the DPU and to find an extremely fast data transmission method for distances in the range of 22-32 km. A strong R&D program should be planned.